\documentclass[%
 reprint,
superscriptaddress,
 amsmath,amssymb,
 aps,
]{revtex4-2}

\usepackage{graphicx}
\usepackage{dcolumn}
\usepackage{bm}

\def\be{\begin{equation}}
\def\ee{\end{equation}}
\def\bea{\begin{eqnarray}}
\def\eea{\end{eqnarray}}

\begin{document}

\preprint{APS/123-QED}

\title{A unified model of dark energy and inflation from the Markov-Mukhanov action}

\author{Hrishikesh Chakrabarty}
\email{hrishikesh.chakrabarty@nu.edu.kz }
\affiliation{Department of Physics, Nazarbayev University, 53 Kabanbay Batyr Avenue, 010000 Astana, Kazakhstan}

\author{Daniele Malafarina}
\email{daniele.malafarina@nu.edu.kz}
\affiliation{Department of Physics, Nazarbayev University, 53 Kabanbay Batyr Avenue, 010000 Astana, Kazakhstan}

\date{\today}

\begin{abstract}
We propose a unified model of dark energy and inflation through the Markov-Mukhanov modification of the Einstein-Hilbert action, where the matter sector is coupled to gravity via a scalar coupling function depending only on the energy density of the matter content. We assume that the coupling function encodes the UV corrections to the standard model of cosmology and we determine the form of the coupling that allows for the dark energy component to be dynamical and act as the inflaton field in the early universe.
Interestingly we show that our model, in order to account for inflation, prefers a dark energy equation of state with $w$ close but not equal to $-1$ in agreement with the latest DESI data.

\end{abstract}

\maketitle


\section{Introduction}\label{sec1}

The standard cosmological model known as $\Lambda$CDM has been very successful in describing the observable Universe while the field of cosmology has seen remarkable advances in the recent decades. These advances are not only a consequence of theoretical developments, but also rely heavily on precise observations such as the cosmic microwave background radiation (CMB) \cite{Planck:2018vyg,Planck:2018jri}, type-IA supernovae (SNIa) \cite{DES:2018paw}, gravitational waves (GW) \cite{LIGOScientific:2019zcs,LIGOScientific:2021aug}, large scale structures \cite{DESI:2025fxa} to name a few. We are living in an era of precision cosmology which allows us to formulate, analyze and potentially discard proposed cosmological models. 

Regardless of this remarkable growth, there are still some open problems that plague the $\Lambda$CDM model. Two of the main issues in modern cosmology are related to the accelerated expansion rate of the Universe. Observations of type-Ia supernovae show that the Universe's expansion is currently accelerating \cite{SupernovaSearchTeam:1998fmf,DES:2018paw}. The unknown cause responsible for this acceleration is called `dark energy' (DE) and numerous models have been proposed to explain the observations while its intrinsic nature remains unknown \cite{Bamba:2012cp}. The simplest form of dark energy is the cosmological constant which accounts for the $\Lambda$ in $\Lambda$CDM and is characterized by an equation of state $p=w\varepsilon$, with $w=-1$. However recent observation seem to favor a dynamical DE with $w$ close but not equal to $-1$ \cite{DESI:2024mwx,DESI:2025zgx}. There also exist alternative models of DE, typically modeled with a scalar field, modifications to the gravitational sector or with braneworld scenarios (see e.g. \cite{Bamba:2012cp} for a detailed review).

Additionally, theoretical models require another period of accelerated expansion in the early universe to solve the horizon and flatness problems. According to the $\Lambda$CDM, without the addition of any extra fields, 
different portions of the CMB should not have been in causal contact when photons were first emitted during the recombination era. However the data shows that the CMB is almost perfectly isotropic with only tiny temperature fluctuations over the whole sky. 
This suggests that the Universe was in thermal equilibrium at the time of emission of the CMB, and thus widely separated portions must have been in causal contact.
Also the spatial geometry of the Universe appears to be very close to flat, while the equations show that any tiny departure from flatness should have grown over time. A phase of accelerated expansion in the early Universe, called `inflation', is typically invoked to explain these observations \cite{Starobinsky:1980te,Guth:1980zm,Linde:1981mu}. The easiest mechanism to drive inflation is given by the introduction of new fields (typically scalar fields) that achieve the desired behavior for the Universe before recombination.
In the last few decades, numerous models of inflation were proposed and many were discarded courtesy to the precision observations. However, there still exist a large array of inflationary models that fit the data and next generation observations are necessary in order to further constrain their validity \cite{Abazajian:2013vfg}. 

Since the physical nature of the scalar fields that drive inflation is unknown attempts have been made at describing both phases of acceleration via a single mechanism. In order to achieve this the model must span an enormous range of energies ($\sim10^{120} {\rm GeV^4}$), ideally via a natural and not fine tuned mechanism involving only one single matter component and/or with minimal modifications to the theory of gravity.

In this article, we aim at building precisely such a model. 
The motivation for such a model comes from the recent DESI results \cite{DESI:2024mwx,DESI:2025zgx} which hint towards a dynamical DE component. While the standard $\Lambda$CDM predicts a constant equation of state for dark energy, the best-fit values from DESI DR I and DR II are found to be slightly above or below $w = -1$ depending on the combination with other external datasets, though remaining statistically consistent with the $\Lambda$CDM model. On the other hand, when the data is confronted with the simplest parametrized dark energy ($w_0w_a$CDM) model \cite{Chevallier:2000qy,Linder:2002et}, the deviation from a constant equation of state of dark energy becomes significant as the results exclude the $\Lambda$CDM limit at $\sim 2\sigma$ to $4\sigma$. This exclusion is driven by a preference for the non-zero evolution parameter, $w_a \neq 0$, which measures the change in $w$ over time at the first order approximation. Other dynamical dark energy models based on scalar fields also show the same preference when tested with DESI data (see e.g. \cite{Cline:2025sbt}). Note that, it still remains crucial to thoroughly examine sources of systematic uncertainties or inconsistencies between the different sets of data that might be contributing to these results. However, these results provide a strong suggestion of departure from the standard $\Lambda$CDM model, thus providing the experimental basis for the investigation of dynamical dark energy models, such as the one proposed here. 

The idea of a unified description of dark energy and inflation is not new. Several models including quintessential inflation \cite{Peebles:1998qn,Uzan:1999ch,Dimopoulos:2002hm,Sami:2004xk,Gonzalez:2004dh,Neupane:2007jm,Park:2024ceu} and interplay of the Higgs boson and the inflaton \cite{Dimopoulos:2018eam} have been proposed using a single scalar field. There are also models based on multiple scalar fields \cite{Masso:2006yk}, entropic cosmology \cite{Cai:2010zw,Cai:2010kp}, the holographic dark energy \cite{Nojiri:2019kkp} and ambitious triple unification of inflation, dark matter and dark energy \cite{Zhang:2021zol,Bose:2008ew,Liddle:2008bm} to name a few.

In this article, we assume that matter is non-trivially coupled to gravity through a scalar coupling function \cite{Markov:1985py} and the universe contains a DE component, whose equation of state parameter is close but not equal to and smaller than $-1$. The non-trivial scalar coupling makes the dark energy component `dynamical' in nature and provides a mechanism for dark energy to matter conversion. When matter and radiation are included in the model, such a universe initially accelerates mimicking an inflationary era. This period ends when enough matter and radiation are created and we enter a decelerating phase. As the universe expands further, it again enters a period of acceleration which mimics the dark energy dominated era. We assume that the equation of state parameter measured today is $w\sim-0.99$ \cite{DESI:2024mwx} and show that the same DE component can cause inflation in the early universe. We verify this claim by testing other values of $w$ against the inflationary constrains on the spectral index of scalar curvature power spectrum and the ratio of tensor-to-scalar perturbations from Planck, BICEP/Keck and Atacama Cosmology Telescope (ACT) observations.

The article is organized as follows: in Sec.~\ref{sec2}, we introduce the basic equations of the theory starting from the action, the equations of motion and the conservation equation. In Sec.~\ref{sec3}, we discuss two possible choices for the free coupling function. In Sec.~\ref{sec4}, we derive the basic equations of cosmology and show how the running constants behave throughout the evolution of the universe. In Sec.~\ref{sec5}, we explore the inflationary consequences of the model deriving the inflationary power spectrum and comparing it with observations. Finally in Sec.~\ref{sec6}, we summarize and results and comment on future directions.    
Throughout the article we use natural units $c = \hbar = 1$ and metric signature $(-,+,+,+)$.

\section{The Markov-Mukhanov action}\label{sec2}

We aim to build a unified model of dark energy and inflation based on two main ingredients:
\begin{itemize}
    \item[(a)] There exist a non-minimal matter-gravity coupling that depends on the energy scale as described by Markov and Mukhanov (MM) in \cite{Markov:1985py}.
    \item[(b)] The exact form of the non-minimal coupling can be expressed via a series of higher order corrections to the matter sector.
\end{itemize}
Then we start with the action \cite{Markov:1985py} 
\begin{equation}\label{eq-action}
    S = \int d^4x \sqrt{-g} \left( \frac{R}{8\pi G_N} + 2\chi (\varepsilon) \mathcal{L}_{\rm m} \right),
\end{equation}
where $R$ is the Ricci scalar, $ \mathcal{L}_{\rm m} $ is the matter Lagrangian. The matter Lagrangian may be chosen at will in order to describe the gravitational source. Here we follow standard prescription for cosmology and adopt a perfect fluid source with $\mathcal{L}_m=\varepsilon$, where $\varepsilon$ is the fluid's energy density while 
the matter-gravity coupling depends only on a scalar function of $\varepsilon$, namely $ \chi(\varepsilon) $. 
One may understand this action as an effective classical description of the departure of a modified gravity theory from General Relativity. Thus this may be viewed as an agnostic approach to departures from GR which does not rely on a given alternative theory but instead aims at determining the properties of such a theory phenomenologically by finding a function $\chi$ that fits observations.
The variation of the action with respect to the metric leads to modified Einstein equation in the form
\begin{equation}\label{eq-ee}
    R_{\mu\nu} - \frac{1}{2}g_{\mu\nu}R = 
   8\pi G_N \tilde{T}_{\mu\nu},
\end{equation}
where
\begin{equation}\label{eq-eff-emt}
    \tilde{T}_{\mu\nu} = \left( \varepsilon\chi \right)_{,\varepsilon}T_{\mu\nu} + (\varepsilon^2 \chi_{,\varepsilon}) g_{\mu\nu},
\end{equation}
is the effective energy-momentum tensor while 
\begin{equation}\label{Tmunu}
    T_{\mu\nu} = \left( \varepsilon + P(\varepsilon) \right)u_\mu u_\nu + P(\varepsilon) g_{\mu\nu},
\end{equation}
is the energy-momentum tensor for a perfect fluid, where $P$ is the fluid's pressure, which we assume may be obtained from an equation of state $P=P(\varepsilon)$, and $u^\mu$ is the four velocity vector field. 
Looking at Eq.~\eqref{eq-ee} and \eqref{eq-eff-emt} we can identify the terms
\begin{equation}
G(\varepsilon) = G_N(\chi \varepsilon)_{,\varepsilon} \; \; \text{and} \; \; \Lambda(\varepsilon) = - 8\pi G_N \varepsilon^2 \chi_{,\varepsilon} 
\end{equation}
as a running gravitational constant $G(\varepsilon)$ and a running cosmological constant $\Lambda(\varepsilon)$. In order to retrieve GR at low densities, for $\chi (\varepsilon)$ we must ensure that as $ \varepsilon \rightarrow 0$ we get $G(\varepsilon) \rightarrow  G_N $.

Since we are writing Einstein's equations with an effective matter source given by $\Tilde{T}^\mu_\nu$, we must require that conservation equations hold for the effective stress-energy tensor, 
i.e.
\begin{equation}
    \nabla_\mu \Tilde{T}^\mu_\nu = 0.
\end{equation}
To see how the effective energy density evolves, we can project it along the four-velocity field $u^\mu$ which gives
\begin{equation}\label{tilde-cons}
    \nabla_\mu \left(\Tilde{\varepsilon}u^\mu \right) + \Tilde{P}\nabla_\mu u^\mu = 0,
\end{equation}
where the quantities with \textit{tilde} are the effective fluid's energy-density and pressure that can be obtained from Eq.~\eqref{eq-eff-emt} as 
\begin{equation} \label{epsilon-tilde}
    \begin{aligned}
        \tilde{\varepsilon} &=  (\chi \varepsilon)_{,\varepsilon}\varepsilon  -\varepsilon^2 \chi_{,\varepsilon} = \varepsilon\chi(\varepsilon), \\
        \tilde{P} &= (\chi \varepsilon)_{,\varepsilon}P+\varepsilon^2 \chi_{,\varepsilon}.
    \end{aligned}
\end{equation}
This allows to define an effective equation of state from $\tilde{P}=\tilde{w}\tilde{\varepsilon}$ as
\be \label{tilde-w}
\tilde{w}=-1+\frac{(\chi \varepsilon)_{,\varepsilon}}{\chi}\left(1+\frac{P}{\varepsilon}\right).
\ee
Notice that if $\chi=1$ we retrieve GR as $G(\varepsilon)=G_N$, $\Lambda(\varepsilon)=0$ and $\tilde{w}=w=P/\varepsilon$.

\subsection{Choice of the free function}\label{sec3}

Our aim is to exchange the introduction of some new matter fields in the early universe with the modifications of the field equations coming from the MM action. Hence, in this formalism, we have the freedom to make a choice for one of the functions $\chi(\varepsilon)$, $G(\varepsilon)$ or $\Lambda(\varepsilon)$. Choosing one of these three functions will fix the other two. 

In order to make such a choice in an agnostic manner (meaning without prior assumptions as derived from some more fundamental theory, as was done for example in \cite{Zholdasbek:2024pxi}) we shall consider generic higher order corrections to the coupling function $\chi(\varepsilon)$ and write it as 
\begin{equation}\label{chi1}
    \chi(\varepsilon) = 1 + \delta (\varepsilon),
\end{equation}
where $\delta(\varepsilon) $ represents departure from minimal coupling. 

Notice that the effective energy density takes the form
\be 
\tilde{\varepsilon}=\varepsilon+\varepsilon\delta(\varepsilon),
\ee 
where the second term can be related to other proposed modifications to the theory in the high curvature regime.

An obvious choice for $ \delta(\varepsilon) $ inspired from Loop Quantum Gravity (LQG) is given by the simple choice $\delta(\varepsilon) = -\varepsilon/\varepsilon_c$.
In LQG, the form of $\delta$ appears from geometric quantum mechanics where the effective equation incorporates the leading corrections from quantum geometry \cite{Ashtekar:2008ay,Willis:2004br,Taveras:2008ke}.
Here the effective density becomes $\tilde{\varepsilon}=\varepsilon-\varepsilon^2/\varepsilon_c$ where $\varepsilon_c$ is a critical energy density at which the corrections become important and typically it is taken of the order of Planck density. 
Such models usually lead to a bounce in the early universe 
as can be seen in Loop Quantum Cosmology (LQC) models \cite{Ashtekar:2006wn,Ashtekar:2008ay,Wilson-Ewing:2012lmx,Agullo:2016tjh} as well as in collapse models \cite{Bambi:2013caa,Chakrabarty:2019omm}. 

However, we note that in the MM formalism the specific expression $\delta(\varepsilon) = -\varepsilon/\varepsilon_c$ may represents merely the first-order correction within a more general perturbative expansion of the coupling function. Consequently, the departure from minimal coupling, $\delta(\varepsilon) $, may be expressed as a power series in the dimensionless parameter $\varepsilon/\varepsilon_c$
\begin{equation}\label{chi2}
    \delta(\varepsilon) = \sum_{n=1}^\infty C_n \left(\frac{\varepsilon}{\varepsilon_c}\right)^n,
\end{equation}
where $C_n$ are dimensionless coefficients that parametrize the strength of the n-th order correction. This expansion allows for a more complete phenomenological description of the coupling function, potentially capturing a wider range of physical phenomena beyond the leading-order approximation.

Now assuming $C_0 = 1$, we can write $\chi(\varepsilon)$ as
\begin{equation}\label{chi3}
    \chi(\varepsilon) = \sum_{n=0}^\infty C_n \left(\frac{\varepsilon}{\varepsilon_c}\right)^n.
\end{equation}
In the following, we shall consider two cases for Eq.~\eqref{chi3}:
\begin{itemize}
    \item[I.] If we consider only the terms up to $n=1$ and absorb $C_1$ in the cutoff density $\varepsilon_c$ by taking $C_1=-1$, then $\chi(\varepsilon)$ becomes
    \begin{equation}\label{chi-lqg}
        \chi(\varepsilon) = 1 - \frac{\varepsilon}{\varepsilon_c},
    \end{equation}
    leading to a cosmological model inspired by but not identical to those studied in LQC.
    \item[II.] If we set $C_n=(-1)^n$ for all $n$ and consider correction terms at all orders, i.e. up to $n = \infty$ then the series converges for $\varepsilon/\varepsilon_c<1$ and 
    $\chi(\varepsilon)$ is given by
    \begin{equation}\label{chi-hayward}
        \chi(\varepsilon) = \frac{1}{1+\varepsilon/\varepsilon_c},
    \end{equation}
    which is just the sum of the infinite geometric series of $\chi(\varepsilon)$.
\end{itemize}
Both models depend only on one parameter $\varepsilon_c$ related to the UV cutoff of GR. Going forward, we shall explore both of these choices to see if given today's content of the universe they may be used to effectively describe early universe cosmology. 

\section{Cosmology}\label{sec4}

\begin{figure*}
    \begin{center}
        \includegraphics[width=8cm]{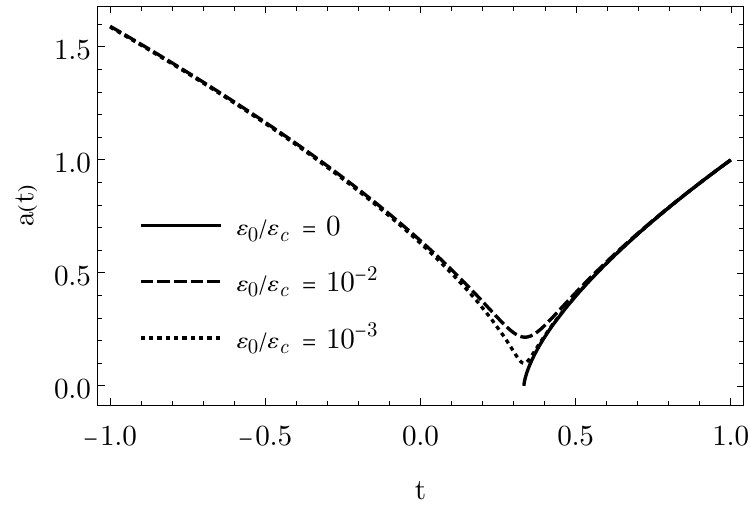}
        \includegraphics[width=8cm]{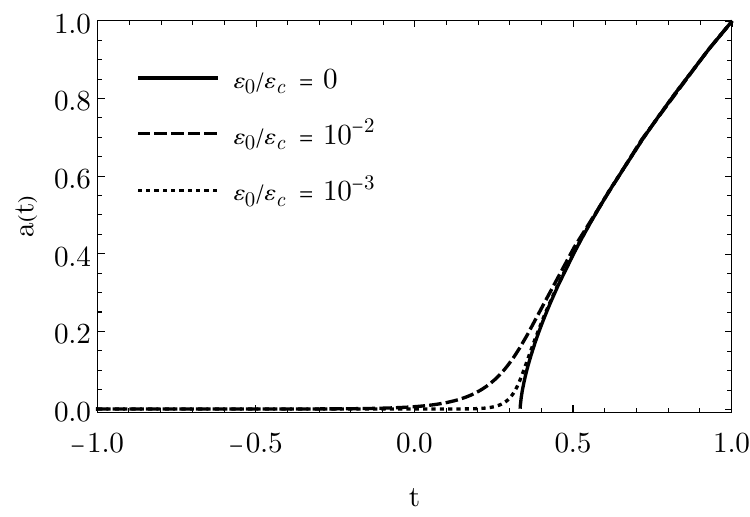}
    \end{center}
    \caption{ Evolution of the scale factor $a$ for a matter dominated cosmology for model I (left panel) and model II (right panel). The left panel shows a bouncing behavior consistent with similar models obtained in LQC. The right panel shows an asymptotically-de Sitter initial state.  
    In both the plots, we have assumed a purely matter dominated universe today. In both plots, the solid line represents a singular dust FRW universe, while the dashed and dotted lines  correspond to different values of $\varepsilon_0/\varepsilon_c$ with values of $\varepsilon_c$ chosen for illustrative purposes. \label{fig-a}}
\end{figure*}

The evolution of the universe is governed by the Friedmann equations. We start by considering a FRW universe with metric given by
\begin{equation}\label{eq-metric}
    ds^2 = -dt^2 + a^2(t)\left(\frac{dr^2}{1-kr^2} + r^2d\Omega^2\right),
\end{equation}
where $a(t)$ is the scale factor, $k$ is the curvature and $d\Omega^2$ is the usual line element on the unit sphere. The Friedmann equations for the MM action are then obtained from the metric \eqref{eq-metric} with the effective energy-momentum tensor \eqref{eq-eff-emt} as
\begin{equation}
    \begin{aligned}
        H^2 &= \left(\frac{\dot{a}}{a}\right)^2 = \frac{8\pi G_N}{3}\varepsilon \chi - 
        \frac{k}{a^2} = \frac{8\pi G_N}{3}\tilde{\varepsilon} - \frac{k}{a^2}, \\
        \frac{\ddot{a}}{a} &= -\frac{4\pi G_N}{3}\left[ (\varepsilon+3P)\chi + 3\varepsilon\frac{\partial \chi}{\partial \varepsilon}\left( \varepsilon + P \right) \right] =\\
        &= -\frac{4\pi G_N}{3} (\tilde{\varepsilon}+3\tilde{P})
    \end{aligned}    
\end{equation}
where $ H = \dot{a}/a $ is the Hubble parameter. The evolution of the effective energy density can be easily obtained from Eq.~\eqref{tilde-cons} as
\begin{equation}
    \dot{\Tilde{\varepsilon}} + 3\frac{\dot{a}}{a}\left(\Tilde{\varepsilon} + \Tilde{P} \right) = 0,
\end{equation}
or in terms of the original fluid quantities as
\begin{equation}\label{eq-conservation}
    \left( \varepsilon \chi \right)_{,\varepsilon}\left[ \dot{\varepsilon} + 3\frac{\dot{a}}{a}\left(\varepsilon + P \right) \right] = 0.
\end{equation}
Since $\left( \varepsilon \chi \right)_{,\varepsilon} \neq 0$ at all times, from the above equation we see that the zero component of the effective energy momentum $\tilde{T}^\mu_\nu$ for an homogeneous perfect fluid is conserved if the classical energy momentum $T^\mu_\nu$ is. 

For a classical fluid with linear equation of state $P=w\varepsilon$, from Eq.~\eqref{tilde-w}, we can define the equation of state of the effective fluid as $\tilde{P} = \tilde{w}\tilde{\varepsilon}$, where $\tilde{w}$ in Eq.~\eqref{tilde-w} can now be written as
\begin{equation}
    \tilde{w} = w + (1+w)\frac{\varepsilon \chi_{,\varepsilon}}{\chi}.
\end{equation}

Let us now explore the two models presented earlier. 
Our goal is to see if, by employing the MM formalism, we can obtain a phase of accelerated expansion in the early universe given only today's matter content for the universe.
For simplicity, we assume a flat universe ($k=0$) which contains only matter ($w_{\rm m}=0$), radiation ($w_{\rm r}=1/3$) and a dark energy component with some equation of state parameter $w_{\rm DE}$. The Friedmann equations can be rewritten in the following way
\begin{align}
    \frac{H^2}{H_{0}^2} &= \sum_i\frac{\Omega_{i0}}{a^{3(1+w_i)}}\chi, \\
    \frac{\ddot{a}}{a} &= -\frac{H_0^2}{2}\sum_i \frac{\Omega_{i0}}{a^{3(1+w_i)}}\chi\Bigg[ 1+ \nonumber \\ & \ \ \ \ \ +3\left( w_i - (1+w_i)\frac{\varepsilon_0}{\varepsilon_c}\frac{\Omega_{i0}}{a^{3(1+w_i)}}\chi \right) \Bigg].
\end{align}
Here $\varepsilon_0 = 3H_0^2/8\pi G_N$ is the energy density of the universe today with $H_0$ being today's Hubble parameter and $\Omega_{i0}$ is today's density parameter for the $i$-th matter component, i.e. with $i=$m, r, DE for matter (m), radiation (r) and dark energy (DE) respectively.

\subsection{Model I}

\begin{figure*}
    \begin{center}
        \includegraphics[width=8.7cm]{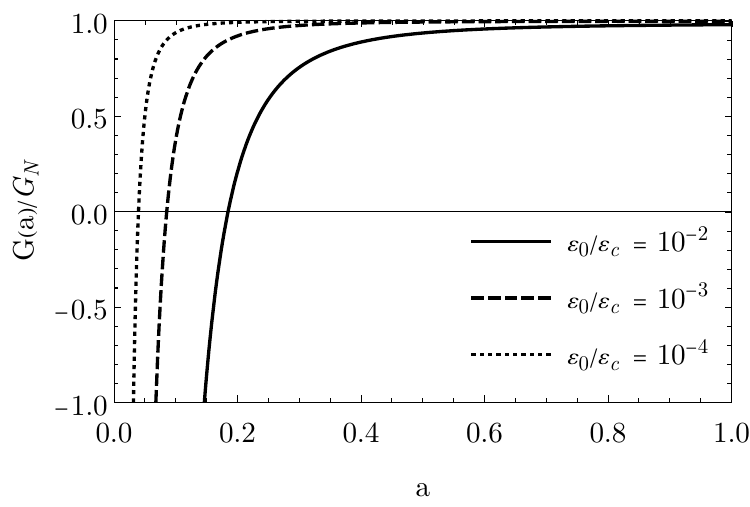}
        \includegraphics[width=8.5cm]{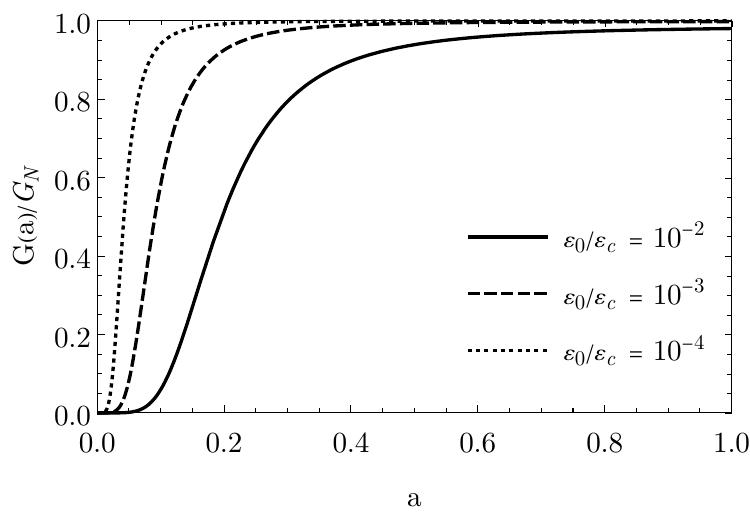}
    \end{center}
    \caption{Behavior of the running Newton's constant $G(\varepsilon)$ as a function of the scale factor for model I (left panel) and model II (right panel). In both the plots, we have assumed a universe with matter, radiation and a dark energy component with $w_{\rm DE}=-0.99$ and all the density parameters are set from the fiducial values obtained in \cite{Planck:2018jri}. The solid, dashed and dotted lines in both the plots correspond to different values of $\varepsilon_0/\varepsilon_c$ with values of $\varepsilon_c$ chosen for illustrative purposes.  \label{fig-g}}
\end{figure*}

The coupling function $\chi(\varepsilon)$ in this case is given by Eq.~\eqref{chi-lqg}.
With this choice, the running Newton's constant and the running cosmological constant are given by
\begin{equation}\label{g1l1}
    \begin{aligned}
        \frac{G(\varepsilon)}{G_N} &= 1 - \frac{2\varepsilon}{\varepsilon_c}, \\
        \frac{\Lambda(\varepsilon)}{8\pi G_N} &= \frac{\varepsilon^2}{\varepsilon_c}.      
    \end{aligned}
\end{equation}
Similarly, we can write the first Friedmann equation as
\begin{equation}
    \left(\frac{\dot{a}}{a}\right)^2 = \frac{8\pi G_N}{3}\varepsilon\left(1 - \frac{\varepsilon}{\varepsilon_c}\right).
\end{equation}
As a first simple example, to illustrate the general behavior of the model, we can solve the Friedmann equation in the case of only matter, $w_{\rm m}=0$, to find $a(t)$. As expected, the resulting solution is of a bouncing kind 
and it is plotted in the left panel of Fig.~\ref{fig-a} with initial condition for the integration taken as $t=1$ today, i.e. $a(1)=a_0=1$. We can see that the model with LQC inspired correction to the coupling function leads to a bounce as one approaches the time for which $\varepsilon$ becomes $\varepsilon_c$. This behavior is consistent with what was found in other LQC models \cite{Ashtekar:2006wn,Ashtekar:2008ay,Singh:2006im}. However, it is important to remark that as a consequence of the MM formalism this model exhibits an induced dark energy component, in contrast with the usual LQC models.

\begin{figure*}
    \begin{center}
        \includegraphics[width=8cm]{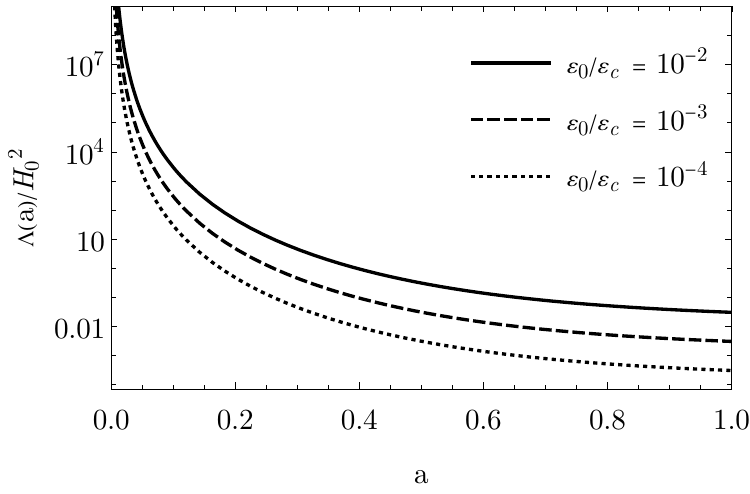}
        \includegraphics[width=8cm]{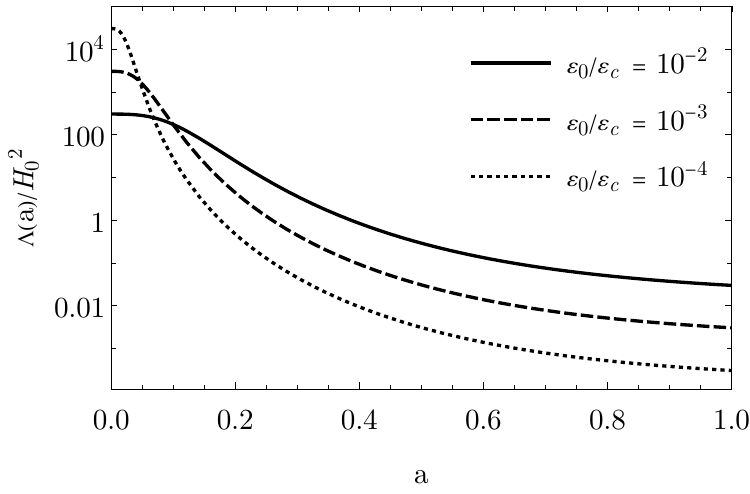}
    \end{center}
    \caption{Behavior of the running cosmological constant $\Lambda(\varepsilon)$ as a function of the scale factor for model I (left panel) and model II (right panel). 
    In both the panels, we have assumed a universe with matter, radiation and a dark energy component with $w_{\rm DE}=-0.99$ and all the density parameters $\Omega_{i0}$ are set from the fiducial values obtained in \cite{Planck:2018jri}. The solid, dashed and dotted lines in both the plots correspond to different values of $\varepsilon_0/\varepsilon_c$ with values of $\varepsilon_c$ chosen for illustrative purposes. \label{fig-l}}
\end{figure*}

We can use the solution of the conservation equation \eqref{eq-conservation} to express $G(\varepsilon)$ and $\Lambda(\varepsilon)$ as functions of the scale factor $a$. 
Considering a fluid made of different components each with linear equation of state parameter $w_i$ we may solve the conservation equation for each component to get $\varepsilon (a) \sim 1/a^{3(1+w_i)}$. Then, since the total density is $\varepsilon=\sum_i \varepsilon_i$, using Eq.~\eqref{g1l1} we get
\begin{equation}
    \begin{aligned}
        \frac{G(a)}{G_N} &= 1-2\frac{\varepsilon_0}{\varepsilon_c}\sum_i\frac{\Omega_{i0}}{a^{3(1+w_i)}}, \\
        \Lambda(a) &= 3H_0^2 \frac{\varepsilon_0}{\varepsilon_c}\left(\sum_i\frac{\Omega_{i0}}{a^{3(1+w_i)}}\right)^2.
    \end{aligned}
\end{equation}
Remember that, we have assumed the scale factor today to be $ a_{0} =1 $. The running Newton's and cosmological constants for Model I are plotted in the left panels of Fig.~\ref{fig-g} and \ref{fig-l} for an universe containing matter, radiation and a DE component with $w_{\rm DE}=-0.99$.
The choice of $w_{\rm DE}$ not being exactly $-1$ is due to the recent DESI BAO results \cite{DESI:2024mwx}, although choosing $w_{\rm DE}=-1$ would lead to qualitatively similar results.
The running Newton's constant in the early universe decreases sharply and becomes negative at scales identified by the critical density parameter signifying the repulsive gravitational force responsible for the bounce. On the other hand, the running cosmological constant increases as  
$a$ becomes smaller. It is important to remember that model I describes a bouncing cosmology and therefore the scale factor can only decrease until it reaches a minimum value determined by the cutoff $\varepsilon_c$. Therefore also $G$ and $\Lambda$ are bounded thus reaching the critical values $G_c=-1$ and $\Lambda_c=8\pi G_N \varepsilon_c$ at the time of the bounce.
 
Finally we plot the comoving Hubble radius $(aH)^{-1}$ as a function of the scale factor in the left panel of Fig.~\ref{fig-hr} for a universe with matter, radiation and a DE component with $w_{\rm DE}=-0.99$. We can see that for the bouncing cosmology the comoving Hubble radius diverges, as expected, as $a$ approaches the bounce.

\subsection{Model II}

In the second case, where the coupling function is given by Eq.~\eqref{chi-hayward},
the running Newton's constant and the running cosmological constant are
\begin{equation}
    \begin{aligned}
        \frac{G(\varepsilon)}{G_N} &= \frac{1}{\left( 1 + \varepsilon/\varepsilon_c \right)^2}, \\
        \frac{\Lambda(\varepsilon)}{8\pi G_N} &= \frac{\varepsilon^2/\varepsilon_c}{\left(1+\varepsilon/\varepsilon_c\right)^2},
    \end{aligned}
\end{equation}
and we can write the first Friedmann equation as
\begin{equation}
    \left(\frac{\dot{a}}{a}\right)^2 = \frac{8\pi G_N}{3}\frac{\varepsilon}{1+\varepsilon/\varepsilon_c}.
\end{equation}

By solving the first Friedmann equation for a simple model with only dust ($w_{\rm m}=0$) we see that the scale factor becomes arbitrarily small in the early universe leading asymptotically to a de Sitter phase as $t\rightarrow -\infty$. 
The solution is plotted in the right panel of Fig.~\ref{fig-a}.

Similar to the previous model, we again assume a perfect fluid matter source composed by several fluids, each with constant equation of state parameter $w_i$. The running Newton's constant and the running cosmological constant become
\begin{equation}
    \begin{aligned}
        \frac{G(a)}{G_N} &= \left[ 1 + \frac{\varepsilon_0}{\varepsilon_c}\sum_i\frac{\Omega_{i0}}{a^{3(1+w_i)}} \right]^{-2}, \\
        \Lambda(a) &= 3H_0^2 \frac{\varepsilon_0}{\varepsilon_c} \left(\frac{\sum_i\frac{\Omega_{i0}}{a^{3(1+w_i)}}}{ 1 + \frac{\varepsilon_0}{\varepsilon_c}\left(\sum_i\frac{\Omega_{i0}}{a^{3(1+w_i)}}\right) }\right)^2.
    \end{aligned}
\end{equation}
In the right panels of Fig.~\ref{fig-g} and \ref{fig-l} we plot these two functions for a universe composed of matter, radiation and Dark Energy with $w_{\rm DE}=-0.99$. Interestingly, in this case, the running $G(a)$ decreases in the early universe but remains positive tending to zero as we approach $a \rightarrow 0$. On the other hand, the running $\Lambda(a)$ approaches a constant value as $a \rightarrow 0$. 
This indicates the weakening of the gravitational interaction in the early universe as it enters an asymptotically  de Sitter phase for $t\rightarrow -\infty$.

\begin{figure*}
    \begin{center}
        \includegraphics[width=8cm]{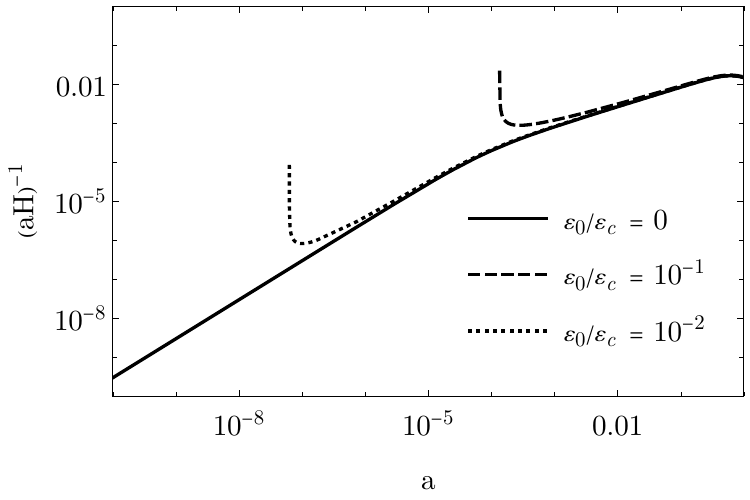}
        \includegraphics[width=8cm]{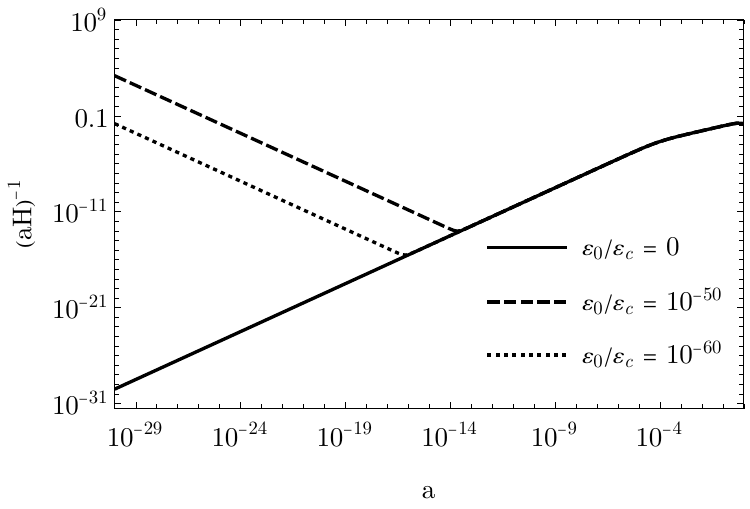}
    \end{center}
    \caption{The comoving Hubble radius as a function of the scale factor for the model with only first order correction to the coupling function (left panel) and the model with infinite orders of correction to the coupling function (right panel). In both the plots, we have assumed a universe with matter, radiation and a dark energy component with $w_{\rm DE}=-0.99$ and all the density parameters are set from the fiducial values obtained in \cite{Planck:2018jri}. The solid line in both the plots correspond to the $\Lambda$CDM model and the dotted and dashed lines correspond to the respective models for different values of $\varepsilon_0/\varepsilon_c$.  \label{fig-hr}}
\end{figure*}

For this case as well, we plot the comoving Hubble radius $(aH)^{-1}$ as a function of the scale factor in the right panel of Fig.~\ref{fig-hr}. As expected, the comoving Hubble radius does not diverge in this case and the behavior can mimic an inflationary phase in the early universe.     

In the next section we shall focus on model II to see the conditions under which a viable inflationary phase in the early universe may be obtained from the MM formalism with $\chi(\varepsilon)$ given by Eq.~\eqref{chi-hayward} without the requirements of additional fields to drive the acceleration.

\section{Inflationary implications for Model II}\label{sec5}

As mentioned, for the second model we have an asymptotically de Sitter initial state of the universe. In this period, the comoving Hubble radius decreases as the universe expands and if there are any primordial fluctuation modes in said period, they would freeze after leaving the horizon. Such modes can act as the seeds for fluctuations in the CMB when they re-enter at a later epoch. In other words, the initial quasi-de Sitter period can mimic an inflationary universe solving the horizon and the flatness problem while also introducing inhomogeneities in the CMB.

\begin{figure*}
    \begin{center}
        \includegraphics[width=7.5cm]{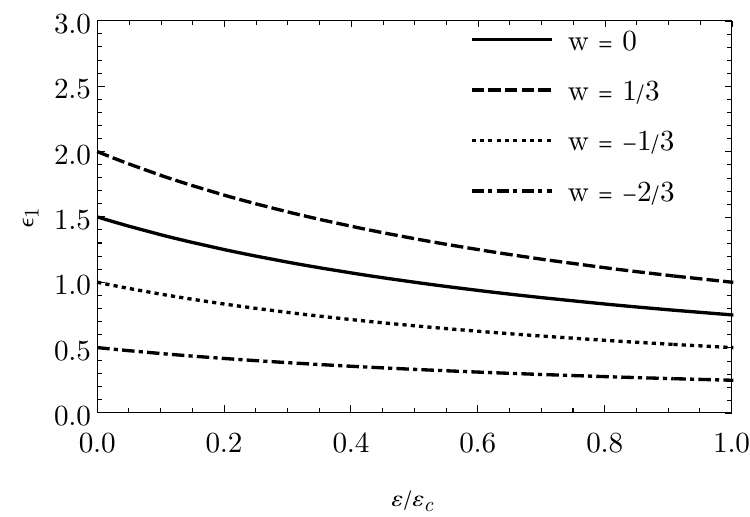}
        \includegraphics[width=7.5cm]{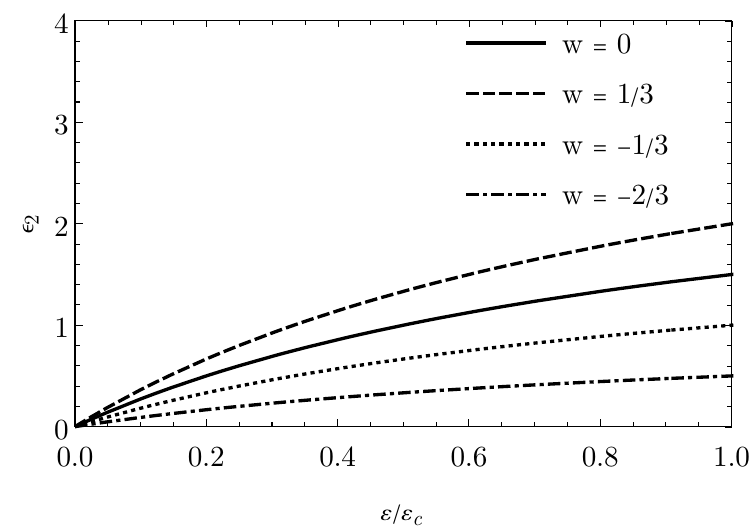}
        \end{center}
    \caption{The slow-roll parameters $\epsilon_1$ (left panel) and $\epsilon_2$ (right panel) as functions of $\varepsilon/\varepsilon_c$ for different matter contents of the universe.  
    Notice that while the first slow-roll parameter tends to zero for $\varepsilon$ large, the second slow roll parameter tends to a constant $\epsilon_2\rightarrow 3(1+w)$ and thus remains small only for suitable values of the equation of state parameter $w$.} \label{fig-sr1}
\end{figure*}

\begin{figure*}
    \begin{center}
        \includegraphics[width=7.5cm]{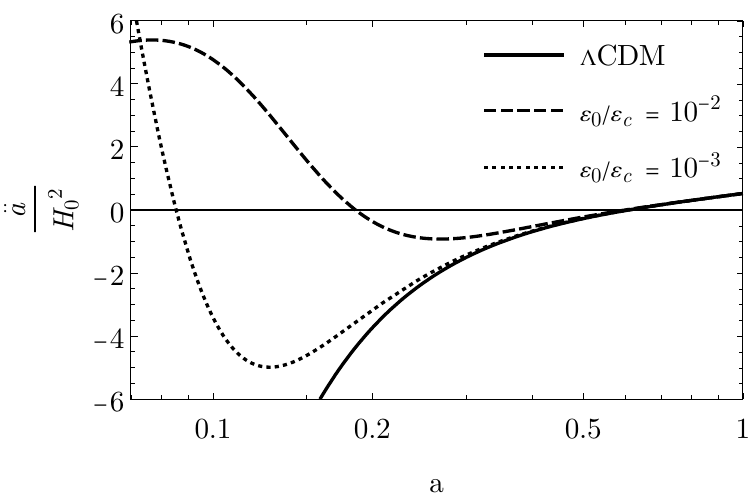}
        \includegraphics[width=7.5cm]{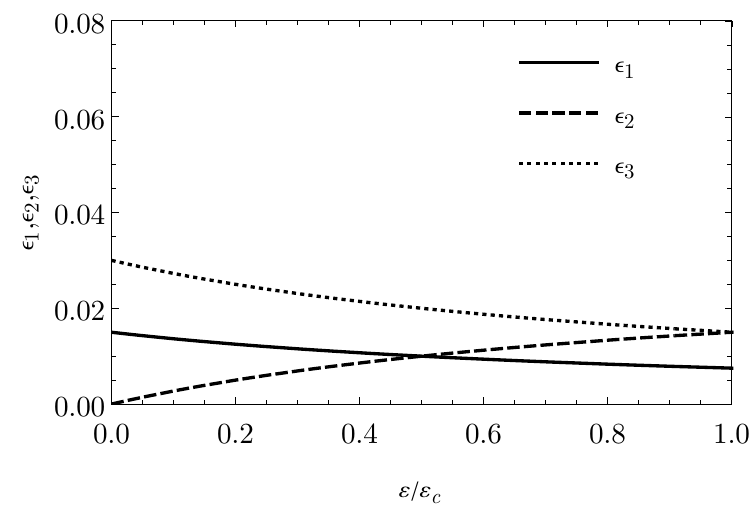}
        \end{center}
    \caption{Left: Acceleration of the universe as a function of the the scale factor. The solid line corresponds to the $\Lambda$CDM model while the dotted and dashed line corresponds to our model II for different values of $\varepsilon_0/\varepsilon_c$. We have assumed a universe with matter, radiation and a dark energy component with $w_{\rm DE}=-0.99$ and all the density parameters are set from the fiducial values obtained in \cite{Planck:2018jri}. Right: The slow-roll parameters $\epsilon_1$, $\epsilon_2$ and $\epsilon_3$ for a dark energy content with equation of state parameter $w_{\rm DE} = -0.99$.  \label{fig-accln-sr}}
\end{figure*}

A viable inflationary model requires the Hubble slow-roll parameters, which characterize the conditions required to sustain the inflationary period, to be smaller than unity.
These slow-roll parameters are defined in terms of the background quantities as
\begin{align}
     \epsilon_1 &= \epsilon = -\frac{\dot{H}}{H^2} , \\
    \epsilon_2 &= \eta = \frac{\dot{\epsilon}}{\epsilon H} , \\
   \epsilon_{n+1} &= \frac{\dot{\epsilon}_n}{\epsilon_n H}, \ \ \ \ n > 1.
\end{align}
To solve the horizon and flatness problems, one requires $\{\epsilon,\eta\} \ll 1$. For our model, the first three of these parameters are
\begin{align}
    \epsilon_1 &= \frac{3}{2}(1+w)\left( \frac{1}{1+\varepsilon/\varepsilon_c}\right), \\
    \epsilon_2 &= 3(1+w)\left( \frac{\varepsilon/\varepsilon_c}{1+\varepsilon/\varepsilon_c} \right), \\
    \epsilon_3 &= 3(1+w)\left( \frac{1}{1+\varepsilon/\varepsilon_c} \right).
\end{align}
In Fig.~\ref{fig-sr1} we plot the first two slow-roll parameters with respect to $\varepsilon/\varepsilon_c$ for different equations of state. 
Both the slow-roll parameters $\epsilon_1$ and $\epsilon_2$ are smaller than unity only for equations of state with $w\leq-1/3$. As can be seen from the right panel of Fig.~\ref{fig-accln-sr}, for sufficiently small values of $w$, i.e. $w\sim - 0.99$, the slow-roll parameters can be small enough for a sustained period of inflation. Therefore, in order to have a successful inflationary early universe in this scenario, one must require a dark energy component with $w_{\rm DE}\simeq - 0.99$  
together with the non-trivial coupling given by the action \eqref{eq-action}. 
If we take the dark energy component to agree with the observed late time acceleration of the universe then
the model may be able to explain 
the early universe acceleration without new exotic fluid components. In addition existing constraints on the behavior of late time acceleration may provide constraints for the early universe inflationary phase, which may in principle be tested.
In the left panel of Fig.~\ref{fig-accln-sr} we plot the acceleration of the universe for this model 
given an energy-momentum composed of matter, radiation and a dark energy component with $w_{\rm DE}\simeq-0.99$. As one can see, the universe initially is accelerating and at some epoch which depends on the scale of correction (dashed line for $\varepsilon_0/\varepsilon_c = 10^{-2}$, and dotted line for $\varepsilon_0/\varepsilon_c = 10^{-3}$) it transitions into a decelerating universe. Later, as the universe evolves we again observe the late-time acceleration phase in accordance with the $\Lambda$CDM. Notice the transition from decelerating to accelerating at late-times is independent of the scale of correction. The solid line in the plot represents the $\Lambda$CDM model.     

\subsection{Linear perturbations and the power spectrum}

To compare the prediction of our model with observations, we need to analyze the model at the perturbative level. Planck observations suggest that the power spectrum of perturbations during inflation was nearly flat \cite{Planck:2018jri}.  Therefore we need to calculate the power spectrum of curvature perturbations generated during the inflationary epoch in the model presented above. Cosmology in the Markov-Mukhanov framework was treated at the perturbative level in the context of asymptotic safety in \cite{Zholdasbek:2024pxi}. Here we closely follow \cite{Chen:2013kta} to obtain the quadratic action, which is second order in the perturbation variables, which we then canonically quantize.

We start from the Markov-Mukhanov Lagrangian in Eq.\eqref{eq-action} which is given by 
\begin{equation}\label{eq-lagr-mult}
        \frac{\mathcal{L}}{\sqrt{-g}} = \mathcal{L}_{\rm G} +\mathcal{\tilde{L}}_{\rm m},
\end{equation}
where the gravity part is $\mathcal{L}_{\rm G}=R/(16\pi G_N)$ and the matter Lagrangian is $\mathcal{\tilde{L}}_{\rm m}=\chi\mathcal{L}_{\rm m}$. We rewrite the matter Lagrangian by adding two Lagrange multipliers as 
\begin{equation} \label{42}
\mathcal{\tilde{L}}_{\rm m} = -\tilde{\rho}(1+e) + \lambda_1 (g_{\mu\nu} u^\mu u^\nu + 1) + \lambda_2 (\tilde{\rho}u^\mu)_{;\mu}.
\end{equation}
Here $\tilde{\rho}=\tilde{\varepsilon}/(1+e)$ is the rest mass density with $e$ being the specific internal energy, $u^\mu$ is the 4-velocity. The Lagrange multipliers $ \lambda_1 $ and $ \lambda_2 $ are for the two constraints: the normalization of the 4-velocity and the conservation of the rest mass density. The variation of the action \eqref{eq-lagr-mult} with respect to the Lagrange multipliers gives the two constraint equations while the variation with respect to the metric gives the Markov-Mukhanov field equations.

Using the standard procedure for splitting the action in the $3+1$ formalism we then construct the quadratic action. The metric in the $3+1$ formalism is
\begin{equation}
    ds^2 = -\alpha^2dt^2 + h_{ij}(dx^i+\beta^idt)(dx^j+\beta^jdt),
\end{equation}
where $\alpha$ and $\beta^i$ are called the lapse function and the shift vector, respectively. A spatially-flat FRW background corresponds to $\alpha = 1$ and $ \beta^i = 0 $. The ADM action can be obtained by putting the above metric in the action 
\begin{equation}
    S = \int dt d^3x \sqrt{h}\alpha(\mathcal{L}_{\rm G} + \mathcal{\tilde{L}}_{\rm m}) ,
\end{equation}
where $h=\det (h_{ij})$. The gravity part can now be written as 
\begin{equation}
        \mathcal{L}_{\rm G} = \frac{1}{16\pi G_N}\left[ ^{(3)}R + \alpha^{-2}\left(K_{ij}K^{ij}- K^2\right) \right],
\end{equation}
where $^{(3)}R$ is the Ricci scalar constructed out of the three-dimensional induced metric $h_{ij}$, $K_{ij}$ is the extrinsic curvature on the constant-time hypersurfaces and $K$ is the trace of $K_{ij}$. The extrinsic curvature is given by
\begin{equation}
    K_{ij} = \frac{1}{2}\dot{h}_{ij} -^{(3)}\nabla_j \beta_i - ^{(3)}\nabla_i \beta_j,
\end{equation}
with $^{(3)}\nabla_i$ representing the covariant derivative with respect to the three dimensional metric $h_{ij}$.  

The variation with respect to lapse and shift (which are Lagrange multipliers of the gravitational system) leads to two additional constraint equations. The four constraint equations can be solved for the Lagrange multipliers and can be used in the action in favor of the perturbation variable(s) of interest.

In the following we are interested in the scalar curvature perturbations $\mathcal{R}$ as the perturbation variable.    
We use the comoving gauge and introduce perturbations to the system in the form
\begin{equation}
    u^{\mu} = ( -1 + u, 0, 0, 0 ), \ \ \ \ \ \ h_{ij} = a^2(t)e^{2\mathcal{R}}\delta_{ij},
\end{equation}
where $u$ is the velocity scalar potential to all orders in perturbations and $\mathcal{R}$ denotes the curvature perturbations in the comoving gauge. 
Similarly we can also perturb the lapse $\alpha$, shift $\beta_i$, Lagrange multipliers $\lambda_n$ ($n=1,2$) and the density field $\tilde{\rho}$ to the first order as
\begin{align}
&\alpha = 1 + \alpha_1, \ \ \ \ \beta^i = \partial^i \beta, \\
& \lambda_n = \bar{\lambda}_n + \delta\lambda_i, \ \ \ \ \tilde{\rho} = \bar{\tilde{\rho}} + \delta\tilde{\rho},
\end{align}
where only the scalar parts of the perturbations are considered. The perturbed version of the continuity and constraint equations can be written using the above perturbation variables. These equations would contain the velocity scalar potential $u$ and the perturbation variable of our choice, namely $\mathcal{R}$. The second order action can now be expressed in terms of these perturbation variables. 
The resulting expression is rather long and
interested readers may check ref.~\cite{Chen:2013kta} (Eq.~(3.23)-(3.26)) for a detailed derivation.
The second order action may be simplified by eliminating the Lagrange multipliers and the other fields in favor of only $\mathcal{R}$. We thus obtain the second order Lagrangian as (see \cite{Chen:2013kta} for details)
\begin{equation}\label{eq-lag1}
    \mathcal{L}_{\mathcal{R}}^{(2)} = \frac{a^3}{8\pi G_N} \left[ \frac{\epsilon_1}{c_e^2}\dot{{\mathcal{R}}}^2 - \frac{\epsilon_1}{a^2}(\partial\mathcal{R})^2 \right],
\end{equation}
where $ c_e $ is the effective or the rest frame sound speed of the dark energy fluid. It is important to point out that no slow-roll approximation has been assumed while deriving this expression, which is general for all fluid inflation models. Notice that the adiabatic sound speed of the fluid $c_s^2 = \dot{\tilde{P}}/\dot{\tilde{\varepsilon}}$ will be negative in the early universe leading to gravitational instability. Hence a source of non-adiabatic pressure fluctuations is required, which can be supplied from the internal degrees of freedom in the dark energy. The presence of the non-adiabatic pressure perturbations then contributes towards the effective sound speed $c_e^2$ which need not be negative (see \cite{Hu:1998kj,Gordon:2004ez} for details).  

Now the second order action with the Lagrangian \eqref{eq-lag1} can be written in the following way
\begin{equation}\label{action-z}
     S^{(2)} = \frac{1}{2} \int d\tau d^3x z^2 \left[ (\mathcal{R}')^2 - c_e^2 (\partial\mathcal{R})^2 \right],  
\end{equation}
where, in order to simplify the notation, we have set $8\pi G_N = 1$ and defined $z^2 = 2a^2\epsilon_1/c_e^2$. Here the primed quantity denotes the derivative with respect to the conformal time $\tau = \int dt/a $. For convenience, we introduce the Mukhanov-Sasaki variable $v = z\mathcal{R}$ \cite{Mukhanov:1988jd,Sasaki:1986hm} and write the action one more time as
\begin{equation}\label{eq-action-ms}
    S^{(2)} = \frac{1}{2} \int d\tau d^3x  \left[ v'^2 - c_e^2 (\partial v)^2 - m(\tau)^2 v^2 \right],
\end{equation}
where 
\bea \nonumber
 m(\tau)^2 &=& -\frac{z''}{z} = (aH)^2 \left[2 - \epsilon_1 + \frac{3\epsilon_2}{2}  -\frac{\dot{c}_e}{c_e H} + \left(\frac{\epsilon_2}{2}\right)^2 +\right. \\ \label{eq-mass}
 & - & \left.  \frac{\epsilon_2(\epsilon_1-\epsilon_3)}{2} - \frac{\dot{c}_e \epsilon_2}{c_e H}  - \frac{\ddot{c}_e}{c_e H^2} + 2\left( 
\frac{\dot{c}_e}{c_e H} \right)^2 \right] ,
\eea 
is a time dependent mass parameter. In the quasi-de Sitter era we have 
\begin{equation}
    aH \simeq -\frac{1}{\tau} (1+\epsilon_1).
\end{equation}
Now the variation of the action \eqref{eq-action-ms} leads to the Mukhanov-Sasaki equation \cite{Sasaki:1986hm,Mukhanov:1988jd},
\begin{equation}
    v'' + c_e^2 \nabla^2 v + m(\tau)^2 v = 0, 
\end{equation}
or in the Fourier basis
\begin{align}\label{eq-ms}
    v''_k + \left(c_e^2k^2 + m(\tau)^2 \right)v_k = 0.
\end{align}
The system can be canonically quantized following the standard procedure and the power spectrum of the quantum fluctuations for the model can be found by solving Eq.~\eqref{eq-ms} with the appropriate boundary conditions. We can write the Mukhanov-Sasaki equation as a Bessel equation by defining $\nu$ from
\begin{equation}\label{eq-nu}
    \tau^2m(\tau)^2 = \nu^2 -1/4,
\end{equation}
and defining a new time variable $T$ as
\begin{equation}
    T = -c_e k \tau ,
\end{equation}
in terms of which Eq.~\eqref{eq-ms} becomes
\begin{equation}\label{51}
    \frac{d^2v_k}{dT^2} + \left[ 1 - \frac{1}{T^2}\left(\nu^2 - \frac{1}{4}\right) \right]v_k = 0.
\end{equation}
All modes exit the Hubble horizon at $ T/c_e = 1 $, i.e. $k = aH$ with sub (super) Hubble scales corresponding to $ T/c_e \gg (\ll) 1 $. On the other hand, the modes cross the sound horizon when $aH = c_e k$ or $T=1$. Now we use another redefinition of the variables setting $ F = v_k/\sqrt{T} $ to write Eq.~\eqref{51} in the final form
\begin{equation}\label{eq-F}
    \frac{d^2F}{dT^2} + \frac{1}{T} \frac{dF}{dT} + \left[ 1 - \frac{\nu^2}{T^2} \right]F = 0.
\end{equation}
This equation can now be easily solved analytically.
The general solution of Eq.~\eqref{eq-F} can be written in terms of Hankel functions as
\begin{equation}
    F(T) = C_1 \mathrm{H}_\nu^{(1)}(T) + C_2 \mathrm{H}_\nu^{(2)}(T),
\end{equation}
where $C_1$ and $C_2$ can be fixed by the boundary conditions. The solution in terms of the Mukhanov-Sasaki variable is
\begin{equation}
    v_k(T) = \sqrt{T}\left[ C_1 \mathrm{H}_\nu^{(1)}(T) + C_2 \mathrm{H}_\nu^{(2)}(T) \right].
\end{equation}
In the limit $c_ek \ll aH$, the Hankel functions take the form
\begin{eqnarray}\label{eq-hankel}
        \mathrm{H}_\nu^{(1)} (T) \Big|_{T \rightarrow 0} &\simeq& \sqrt{\frac{2}{\pi}} e^{-i\frac{\pi}{2}} 2^{\nu - \frac{3}{2}} \frac{\Gamma (\nu)}{\Gamma(\frac{3}{2})} T^{-\nu}, \\ 
        \mathrm{H}_\nu^{(2)} (T) \Big|_{T \rightarrow 0} &\simeq& -\sqrt{\frac{2}{\pi}} e^{-i\frac{\pi}{2}} 2^{\nu - \frac{3}{2}} \frac{\Gamma (\nu)}{\Gamma(\frac{3}{2})} T^{-\nu}.
\end{eqnarray}
We use the Bunch-Davies conditions \cite{Bunch:1978yq} to find $C_1$ and $C_2$. For the mode functions deep in the sub-Hubble limit, i.e. $c_ek\gg aH$ (or $T\rightarrow \infty$), 
\begin{equation}
    v_k(T) \Big|_{T\rightarrow \infty} \rightarrow \frac{1}{\sqrt{2c_ek}} e^{iT} = \sqrt{T} C_1 \mathrm{H}_\nu^{(1)} (T)\Big|_{T \rightarrow \infty}.
\end{equation}
This yields
\begin{equation}
    C_1 = \frac{1}{\sqrt{2c_ek}} \sqrt{\frac{\pi}{2}} e^{i\left( \nu + \frac{1}{2} \right)\frac{\pi}{2}}, \ \ \ \ C_2 = 0. 
\end{equation}
Now the final expression for the mode functions becomes
\begin{equation}\label{eq-vk-sol}
v_k(T) =\frac{ e^{i\left( \nu + \frac{1}{2} \right)\frac{\pi}{2}} }{2} \sqrt{\frac{\pi T}{c_ek}}  \mathrm{H}_\nu^{(1)} (T).     
\end{equation}
Finally, the power spectrum of comoving curvature perturbations in our model can be written as
\begin{equation}
    \mathcal{P}_{\mathcal{R}}(k) = \frac{k^3}{2\pi^2}|\mathcal{R}_k|^2 = \frac{k^3}{2\pi^2} \frac{|v_k|^2}{z^2}, 
\end{equation}
where, $v_k$ is given by Eq.~\eqref{eq-vk-sol}. Using the expression for the Hankel function in the super-Hubble limit from from Eq.~\eqref{eq-hankel}, the power spectrum becomes
\begin{equation}
    \mathcal{P}_{\mathcal{R}}(k) = \frac{2^{2\nu-3}}{8\pi^2\epsilon_1 c_e} \left( \frac{\Gamma(\nu)}{\Gamma(\frac{3}{2})} \right)^2  \left(\frac{H}{m_{\rm pl}}\right)^2 \left( \frac{c_e k}{aH} \right)^{3-2\nu},
\end{equation}
where $a$, $H$ and $\epsilon_1$ are evaluated at the sound horizon exit and we have reinstated the factor of $ m_{\rm pl}^2 = 1/8\pi G_N$. Now, we can extract the spectral index of the scalar perturbations 
\begin{equation}\label{eq-ns}
n_s - 1 \simeq 3 - 2\nu,     
\end{equation}
where $\nu$ is defined from Eq.~\eqref{eq-nu} and it is a function of the slow-roll parameters, the effective sound speed and its derivatives. It can easily be verified that for sufficiently small values of slow-roll parameters and a nearly constant effective sound speed, the power spectrum of scalar curvature fluctuations is flat. 

Finally the tensor power spectrum $\mathcal{P}_{T}$ is obtained in exactly the same manner as the single scalar field slow-roll case. We obtain 
\begin{equation}
    \mathcal{P}_{T} = \frac{2H^2}{\pi^2 m_{\rm pl}^2},
\end{equation}
so that the tensor-to-scalar ratio $r$ is given by
\begin{equation}\label{eq-r}
    r \simeq 16\epsilon_1 c_e.
\end{equation}

\subsection{Comparing with observations}

\begin{figure*}
    \begin{center}
        \includegraphics[width=7.5cm]{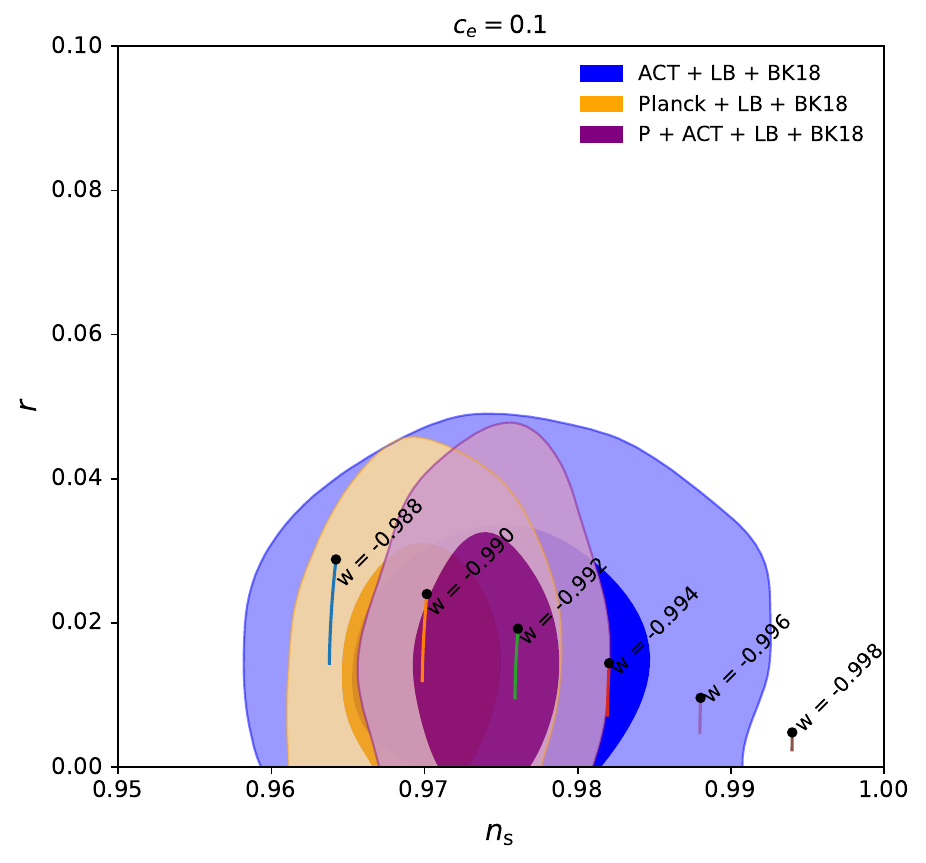}
        \includegraphics[width=7.5cm]{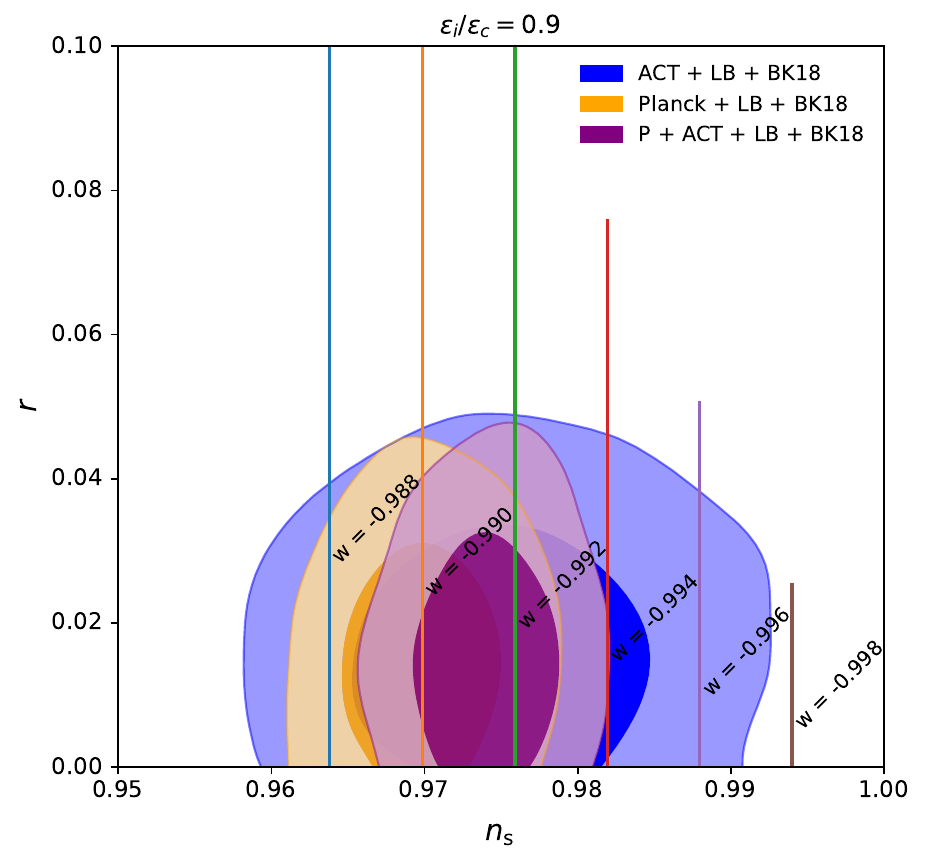}
        \end{center}
    \caption{Comparison between model prediction and observational data. The blue contours indicates 1$\sigma$ and 2$\sigma$ confidence region on $n_s$ and $r$ from ACT + LB + BK18 data, the orange contours indicates 1$\sigma$ and 2$\sigma$ confidence region on $n_s$ and $r$ from Planck + LB + BK18 data and the purple contours indicates 1$\sigma$ and 2$\sigma$ confidence region on $n_s$ and $r$ from Planck + ACT + LB + BK18 data (see text). The vertical lines in both panels correspond to model prediction value of $n_s$ and $r$ for different equation of state parameters of dark energy. Left panel: $\varepsilon_i/\varepsilon_c$ varies along the vertical line while $c_e = 0.1$ is fixed. The black dots represent the requirement of 60 e-folds of inflation. Right panel: The effective sound speed $c_e$ varies along the lines decreasing downwards while $\varepsilon_i/\varepsilon_c = 0.9$ is fixed.  \label{fig-nsr}}
\end{figure*}

We aim now to test the inflationary prediction of our model with observational data from Planck \cite{Planck:2018jri}, BICEP/Keck \cite{BICEP:2021xfz}, Atacama Cosmology Telescope (ACT) \cite{ACT:2025fju,ACT:2025tim} and Dark Energy Spectroscopic instrument (DESI) baryon acoustic oscillation (BAO) \cite{DESI:2024lzq,DESI:2024mwx,DESI:2024uvr}. Specifically, we consider the observational constraints on the tensor-to-scalar ratio $r$ based on BICEP/Keck (BK18) data and constraints on the scalar spectral index $n_s$ driven by Planck, ACT and a combination of Planck and ACT (Planck + ACT) datasets. The combined dataset also includes CMB lensing and BAO (LB) in all cases. The three combinations we consider are named as follows: (a) ACT + LB + BK18, (b) Planck + LB + BK18 and (c) Planck + ACT + LB + BK18. These sets of data provide some of the strongest constraints on the cosmological parameters and have been able to rule out a number of inflationary models in the past. However, note that the ACT and DESI data should be taken with a grain of salt because of a reported degeneracy between BAO data from DESI under the assumption of the standard cosmological model and the CMB data \cite{Ferreira:2025lrd}, though the difference in the most cases is just about $2\sigma$.

Recent DESI results hint towards a dynamical nature of dark energy instead of a cosmological constant $\Lambda$  which has an equation of state, $w_{\Lambda} = -1$ \cite{DESI:2024mwx}. Analysis of DESI BAO data with $w$CDM model constrains the equation of state to $w_{\rm DE} = -0.99^{+0.15}_{-0.13}$. The constraints become stronger when CMB and PantheonPlus supernovae samples are included $w_{\rm DE} = - 0.997 \pm 0.025$. In our analysis, we assume a universe containing a dark energy component with the current equation of state close to the value constrained by DESI with the $w$CDM model. This dark energy component gains a dynamical nature by virtue of the Markov-Mukhanov action and, as discussed previously, can be modeled as the driver of a period of accelerated expansion in the early universe, depending on the choice of the matter-gravity coupling $\chi$.

Now we use the slow-roll parameters, as evaluated for our model II, to calculate $n_s$ and $r$ through Eqs.~\eqref{eq-ns}, \eqref{eq-nu} and \eqref{eq-mass} and for simplicity we assume a constant/slowly-varying effective speed of sound (i.e. $c_e \simeq {\rm const.}$), which is a reasonable assumption in the quasi-de Sitter phase. In Fig.~\ref{fig-nsr}, we show the marginalized contours for $n_s$ and $r$ for the datasets considered along with the predictions of our model. Vertical lines in both the panels of the figure refer to different values of the equation of state parameter in the DESI BAO bound. In the left panel of Fig.~\ref{fig-nsr} we have fixed the effective sound speed $c_e = 0.1$ while $\varepsilon_i/\varepsilon_c$ varies along the line (increasing downwards). Here $\varepsilon_i$ is the energy density of the universe at the sound horizon exit. The ratio $\varepsilon_i/\varepsilon_c$ can also be expressed in terms of the number of e-folds of inflation $N$, written as
\begin{align} \label{N}
    N = \ln \left(\frac{a_e}{a_i}\right) = \ln \left( \frac{\varepsilon_i}{\varepsilon_e} \right)^{1/3(1+w)},
\end{align}
where $a_i$ and $a_e$ are the scale factors at the beginning and end of inflation respectively, $\varepsilon_e$ is the energy density at the end of inflation and we have used the continuity equation in deriving the above relation. 
Hence the ratio can be written as 
\begin{align}
    \frac{\varepsilon_i}{\varepsilon_c} = \frac{\varepsilon_e}{\varepsilon_c} e^{3(1+w)N},
\end{align}
and the slow roll parameters may also be expressed in terms of $N$ as
\begin{align}
     \epsilon &=  -\frac{d\ln H}{dN}, \\ 
     \eta &= \frac{|d \ln \epsilon|}{dN} .
\end{align}
The black dot on every line in the left panel of Fig.~\ref{fig-nsr} refers to the energy density needed at the horizon exit in order to have about 60 e-folds of inflation assuming that the inflationary period ends just before reheating ($T_{\rm reh} \simeq 1 {\rm MeV}$). In the right panel, we have fixed $\varepsilon_i/\varepsilon_c = 0.9$ and $c_e$ varies along the vertical lines (decreasing downwards). As we can see, models with $w \simeq -0.99$ and smaller values of the effective sound speed $c_e$ agree well with the recent observational data. On the other hand a model with $w_\Lambda=-1$ appears to be excluded.

\section{Summary}\label{sec6}

We considered a cosmological model obtained from a modification of the field equations of General Relativity as proposed by Markov and Mukhanov in \cite{Markov:1985py}. 
For the matter content of the universe we considered only matter, radiation and a dark energy component in accordance with recent observation. We then obtained a phase of accelerated expansion in the early universe via a suitable and natural choice of the 
matter-gravity coupling. In this framework the role of the inflaton field is replaced by the effects of the matter-gravity coupling on the universe's content.
The most noteworthy result of our analysis is that in order for the model to match the observational data from the CMB we must require that the late universe dark energy equation of state departs slightly from the cosmological constant. This is in accordance with recent observations by DESI that suggest a dynamical nature for dark energy. In our model, the dynamical nature of dark energy is achieved through the matter gravity coupling.
If confirmed, the connection of such dynamical nature of dark energy with a connection to the matter-gravity coupling may turn out to have important implications for our understanding of gravity at high energies and at large scales.

\section*{Acknowledgement}
DM and HC acknowledge support from Nazarbayev University Faculty Development Competitive Research Grant Program No. 040225FD4737. HC would like to thank Seong Chan Park for insightful discussions and Yukawa Institute for Theoretical Physics, Kyoto for warm hospitality where part of this work was initiated.

\bibliographystyle{apsrev}
\bibstyle{apsrev}
\bibliography{ref}

\end{document}